\documentclass[prb,twocolumn,showpacs,showkeys,amsmath,amssymb]{revtex4}

\usepackage{graphicx}
\usepackage{dcolumn}
\usepackage{bm}
\usepackage{courier}

\begin{document}



\title{Geometrical Probability Distribution Functions for Cable-in-Conduit Conductors with Simply and Multiply Connected Cross-Sections}
\author{A.Anghel}
\affiliation{Paul Scherrer Institute, CH-5232
Villigen PSI, Switzerland}

\date{\today}

\begin{abstract}
A geometrical method is presented for the calculation of the strand distribution functions for cable-in-conduit
superconductors with simple and multiply connected cross-sections. The method is
illustrated on  different cable designs with simply and multiply connected structures. We start with the simple case of a round cable (a simply connected structure) and continue with some multiply connected structures: a cable with a central channel
then one  with a central channel and wrapped petals and finally a cable with segregated copper and central channel. Analytical
relations are given which can be used in numerical simulations to calculate average values of certain cable properties like the average electric field.
\end{abstract}

\pacs{23.23.+x,56.56.Dy}
\keywords{probability, distribution, geometrical, superconducting cable}
\email{anghel@psi.ch}

\maketitle  

\section{Introduction}
\begin{figure}[tb]
	\centering
		\includegraphics[width=6cm,keepaspectratio]{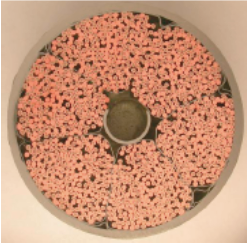}
	\caption{\label{fig:cicc} Color online. Typical Cable in Conduit Conductor with central channel and inter-petal void spaces.}
\end{figure}

A cable-in-conduit conductor (CICC) is usual made of a bunch of
strands encapsulated in a metal conduit. The bunch consists of a
large number of strands, around 1000, twisted together in many
different stages, starting from a simple basis unit. Usually this
is a triplet, a twist of three strands, but cables where the basic
unit is a more complicated build such as a copper core surrounded
by a number of superconducting strand twisted around it or even a
relatively large braid are known. Not all strands in the cable are
or should be by necessity superconducting. Segregated copper
strands ca replace some of the superconducting ones either at the
level of the basic unit or at higher stages. For example, instead
of a full superconducting triplet one can have a triplet with the
structure 2S+1Cu consisting of two superconducting and one copper
wires. Alternatively, copper segregation can be produced by
inserting copper wires or copper subcables in the cabling process
at certain upper stages. The cross-section of the cable can be
therefore very rich in structure showing either a dispersed
segregation of copper wires if copper was introduced at the basic
stage or islands of copper if copper is introduced at later
stages. A typical cable in conduit conductor is shown in Fig.\ref{fig:cicc}

Another source of inhomogeneity in the cable cross-section is related to the
void spaces. Some cables are provided with a central channel added in order
to improve the helium circulation and to reduce the pressure drop. In other
cables the last stage (called ``petal'' for obvious reasons) is wrapped with
a steel tape creating additional inter-petal spiral channels. In this respect we speak from cables which have a simply connected or a multiply connected \footnote{Informally, a thick object in our space is simply connected if it consists of one piece and does not have any "holes" that pass all the way through it. For example, neither a doughnut nor a coffee cup (with handle) is simply connected, but the surface of a hollow rubber ball is simply connected. In two dimensions, a disk is simply connected but a disk with holes is not. Spaces that are connected but not simply connected are called non–simply connected or, in a somewhat old-fashioned term, multiply connected.(Adapted from \textit{http://en.wikipedia.org/wiki/Simply\_connected\_space})} structure. 

Due to the self-field effect, the magnetic field distribution in
the cable cross-section is not uniform. Through the dependence of
critical current on field, the cable will show a non-uniform
distribution of electric field in the cross-section if the
volt-ampere characteristic of the strands is of power-law type.
The measured electrical field(using voltage taps on the cable jacket) is a kind of average over an
ensemble of strands. The calculation of the average electric field
in a cable is based on the ergodic hypothesis \cite{ref1} which
states that all strands trajectories are equivalent and the length
average of one strand can be replaced by an ensemble average. This
is similar to the principle from the statistical mechanics where the time average of a
physical quantity can be replaced by an ensemble average. It was
demonstrated in \cite{ref2} that for short length the ergodic
hypothesis is not true and that in this case a statistical
interpretation of the experimental results based on the central
limit theorem should be made. For long cables sections however, with length
of hundreds of the last stage twist pitch it seems, based on
numerical simulation of strand trajectories, that the ergodic
hypothesis holds \cite{ref2}. In this case the ensemble average
can be calculated using a geometric probability distribution
function as applied for the first time by Ekin \cite{ref3} for
filaments in a strand and extended later in \cite{ref4} for
simple connected full-size CICC. However, no mathematical proof exists for the time being that the
twisted cables are ergodic for any number of stages and any combination of twist pitches.

According to Ekin's proposal\cite{ref3}, the average electric field (or any other
observable) is calculated using the formula

\begin{equation}
\label{eq1} \left\langle E \right\rangle \overset{\wedge}{=}\int
{E\left( r \right)w\left( r \right)dr=\int {E\frac{dA_E}{A}} }
\end{equation}

where $E\left( r \right)$ is the local electric field and $w\left(
r \right)dr$ the probability density that a strand at position $r$ senses an
electrical field $E(r)$

\begin{equation}
\label{eq2}
w\left( r \right)dr\overset{\wedge}{=}\frac{dN_E }{N}
\end{equation}

with $dN_E $ the number of strands sensing the same electric field
$E$ and $N$ is the total number of strands.

The last equality in Eq.(\ref{eq1}) is based on the important
assumption that the strands are  distributed in the cable cross
section such that a density (number of strands per unit area)
can be defined. We assume therefore that we always can write

\begin{equation}
\label{eq3}
\begin{array}{c}
 dN_E =\rho dA_E \\ \\
 N=\rho A \\
 \end{array}
\end{equation}

In the above equation, $\rho $ is the density of strands in the
cable cross-section, $A$ is the total cable cross-sectional area
and $dA_E $ represents a geometrical area element with the
property that inside this area the electric field is constant. In
other words it is the set of cable cross-section points having the
same electrical field $E$

\begin{equation}
\label{eq4}
dA_E \overset{\wedge}{=}\left\{ {r\in A\left| {E\left( r \right)=E} \right.}
\right\}
\end{equation}

In most of the cases of interest the density $\rho$ is constant and we we
limit ourselves here to this case. Generalization to non uniform
distribution of strands is straightforward and will be developed in the next sections. At this point it is
important to make the difference between uniform density of
strands and random strand position. A uniform density of strands
in the cable cross section can be obtained also with strands
twisted in concentric layers much similar to the superconducting
filaments in the strand itself.Unfortunately this arrangement is
not ergodic. By contrast, in multiple twisted cables the
strands are much more randomly distributed as a consequence of the
erratic overlapping of multiple twist stages with different
(incommensurate) twist pitches. Therefore, it is generally
believed that such cables satisfy the ergodicity condition.
Therefore for the definition of the geometric probabilities we
request first the randomness and then the uniform density. The
randomness and ergodicity are the necessary conditions to
calculate averages the way defined in Eq.(\ref{eq1}).

\begin{figure}[tb]
\centerline{\includegraphics[width=8cm,keepaspectratio]{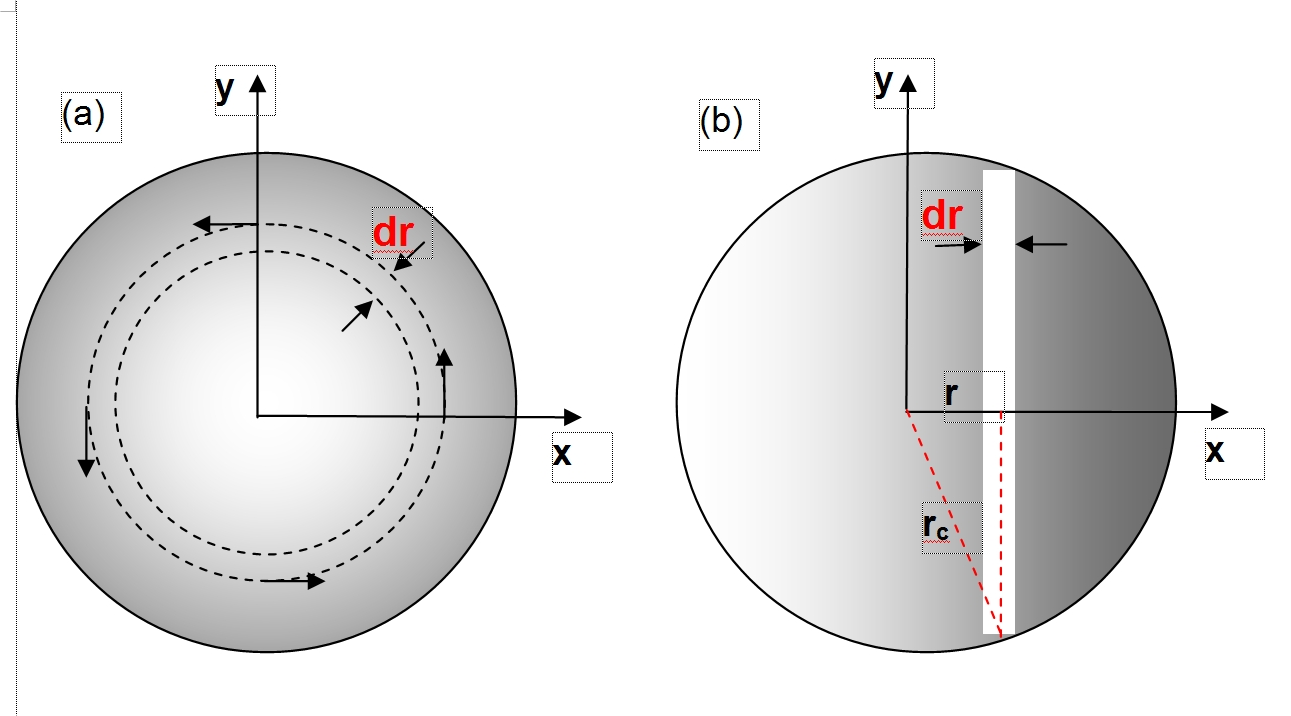}}
\caption{\label{fig1} Color on line. Typical elemental areas for geometrical
probability. a) no external field, self field has a radial distribution, b) superposition of an external uniform field (along Oy-axis) with the self-field of the cable results in a linear field distribution (field gradient) along the Ox-axis.
Dark gray corresponds to higher  magnetic field.}
\end{figure}

The form of the elemental area can change from case to case. For
the self-field problems it is a linear stripe perpendicular to the
magnetic field gradient. For a round cable with transport current
and no external field, the elemental area is a circular ring. The
form of $dA_E $ should be therefore defined for each problem at
hand but because in this paper we will discuss only the self-field
effect the elemental area will be always a stripe as shown in
Fig.\ref{fig1}b.

\section{Illustrative calculation for a simply connected cable contour} 
The simplest case of a simply connected cable is a
circular cable without internal
voids. As discussed above, the superposition of the uniform
external field with the azimuthal field of the cable results in a
linear variation of the magnetic over the cable cross-section. In
this case the elemental area is a stripe (Fig.\ref{fig1}b), parallel to the external field direction with
area given by

\begin{equation}
\label{eq5}
dA_E =dA\left( r \right)=2\sqrt {r_c^2 -r^2} dr
\end{equation}

where $r_c$ is the cable radius and $r$ is the coordinate along
the field gradient (Ox-axis in our case). The total cable cross-section area is

\begin{equation}
\label{eq6}
A=\pi r_c^2
\end{equation}

and using Eqs.(\ref{eq2}) and (\ref{eq3}) we get the following
expression for the probability distribution function

\begin{equation}
\label{eq7}
w_0\left( r \right)=\frac{2\sqrt {r_c^2 -r^2} }{\pi r_c^2 }
\end{equation}

\begin{figure}[tb]
\includegraphics[width=8cm,keepaspectratio]{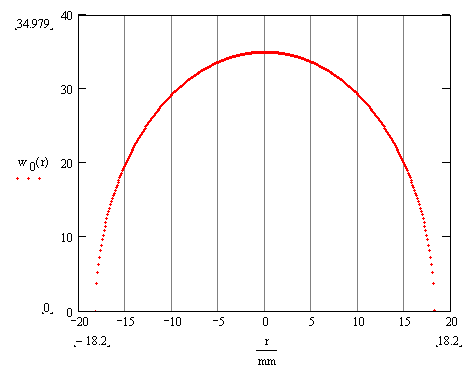}
\includegraphics[width=8cm, keepaspectratio]{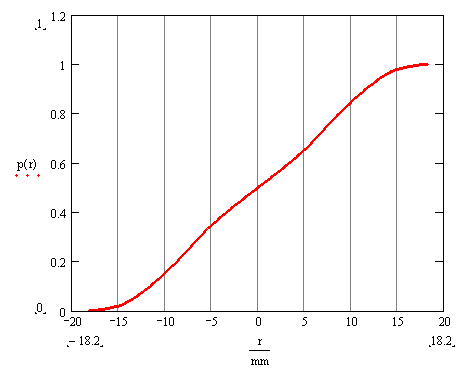}
\caption{\label{fig2} Color online. The probability distribution function $w_0(r)$(left panel) and
the probability $p_0(r)$ (right panel), for a cable with a simple connected contour.}
\end{figure}

This is a strongly nonlinear function with a large slope close to
$r=-r_c$ and $r=r_c$ and is represented graphically in
Fig.\ref{fig2}. Further, it can be checked by direct integration
that $w_0\left( r \right)$, Eq.(\ref{eq7}) satisfies the conditions
of a probability distribution function i.e. we have first,

\begin{equation}
\label{eq8}
\int\limits_{-r_c }^{r_c } {w_0\left( r \right)dr=1}
\end{equation}

and second, the probability that a given strand is placed
somewhere between $-r_c $ and $r$ is smaller than 1 for any $r\in[-r_c, r_c]$

\begin{equation}
\label{eq9}
p_0\left( r \right)=\int\limits_{-r_c }^r {w_0\left( r \right)dr\leqslant 1}
\end{equation}

The infinitesimal probability $dp_0$ is by definition

\begin{equation}
\label{eq10}
dp_0\left( r \right)\overset{\wedge}{=}w_0\left( r \right)dr=\frac{2\sqrt{r_c^2-r^2}}{\pi r_c^2}dr
\end{equation}

\begin{figure}[tb]
\centerline{\includegraphics[width=8cm,keepaspectratio]{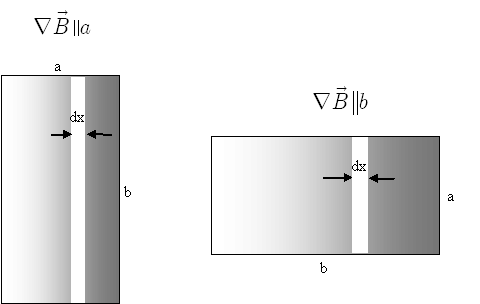}}
\caption{\label{fig3} Geometrical parameters for rectangular
conductor of width $a$ and height $b$. Left panel $\nabla B||a$, right panel $\nabla B||b$. }
\end{figure}

Similarly, for a rectangular conductor of width $a$ and height $b$ as in Fig.\ref{fig3} we obtain

\begin{equation}
\label{eq11}
\begin{array}{l}
 w_a\left( x \right)dx=\dfrac{bdx}{ab}=\dfrac{1}{a}dx\mbox{ \quad \quad for }\nabla B\left\|a \right. \\ \\
 w_b\left( x \right)dx=\dfrac{adx}{ab}=\dfrac{1}{b}dy\mbox{ \quad \quad for }\nabla B\left\|b \right. \\
 \end{array}
\end{equation}

In this case the elemental area is either $dA_E =b~dx$ or $dA_E
=a~dx$ and the total area is $A=ab$. The probability distributions
are both uniform but can differ substantially because $a<b$.

\section{General problem, multiple-connected cable contours}
Multiple connected cables have holes or regions in the
cross-sectional area occupied by another material and not by the
superconducting strands. The holes could be for example the central
channel or the inter-petal spaces for the conductors with wrapped
last stage. Non homogeneous regions could be the copper
islands in case of segregated copper cables. The holes and the
inhomogeneous regions can be fixed ( they do not change the position in 
the cable cross-section as one moves along the cable axis) or itinerant 
( they have variable position in the cable cross-section as one moves along the cable axis). The central
channel is a good example of a fixed hole while the inter-petal
spaces are itinerant holes. They rotate as part of the last stage
with the same twist pitch as the cable's last stage. As we will
see later the area contribution of fixed holes and itinerant holes
to the probability distribution function is quite different. The
same holds for copper islands, although fixed copper islands are
not found in the modern cable design.

\begin{figure}[tb]
\centerline{\includegraphics[width=6cm,keepaspectratio]{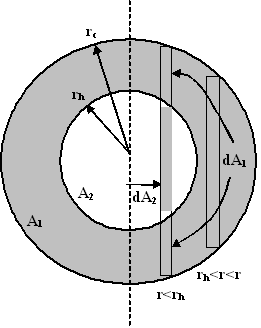}}
\caption{\label{fig4} Geometry for a multiple connected cable}
\end{figure}

In order to illustrate the calculation method for multiple
connected cables let as refer to the geometry presented in
Fig.\ref{fig4}. The cable cross-section consists of two regions
$A_1 $ and $A_2 $ with the property: $A=A_1 \cup A_2 ;\mbox{ }A_1
\cap A_2 =\emptyset$, where $\emptyset$ is the empty set. Transversally, there is a magnetic field gradient 
created by the superposition of the background field with the self field 
of the cable as shown in Fig.\ref{fig1}.

For $|r|<r_h$ the infinitesimal probability is by definition

\begin{equation}
\label{eq12} w_{r<r_h}\left( r \right)dr=\frac{dN_1 +dN_2 }{N}
\end{equation}

with $dN_i , \quad i=1,2$ the number of strands in the
infinitesimal area $dA_i $. Assuming a uniform strand distribution 
in the two regions with densities $\rho_1$ and $\rho_2$ we have

\begin{equation}
\label{eq13}
\begin{array}{l}
 dN_1 =\rho _1 dA_1 ;\mbox{ }dN_2 =\rho _2 dA_2 ; \\ \\
 N=\rho _1 A_1 +\rho _2 A_2 ;\mbox{ }A=A_1 +A_2
 \end{array}
\end{equation}

Substituting this back in Eq.\ref{eq12} we obtain

\begin{equation}
\label{eq14}
\begin{array}{c}
 w_{r<r_h}\left( r \right)dr=\dfrac{\rho _1 dA_1 +\rho _2 dA_2 }{\rho _1 A_1 +\rho _2 A_2 }=\\ \\
=\dfrac{\rho _1 }{\rho _1 A_1 +\rho _2 A_2 }dA_1 +\dfrac{\rho_2 }{\rho_1 A_1 +\rho _2 A_2 }dA_2 = \\ \\
=\dfrac{A}{A_1 +\lambda A_2 }\dfrac{dA_1 }{A}+\dfrac{\lambda A}{A_1 +\lambda A_2 }\dfrac{dA_2 }{A}
\end{array}
\end{equation}

with $\lambda ={\rho _2 }/{\rho _1 }$.

For $r_h<|r|<r_c$ we have similarly

\begin{equation}
\label{eq15} 
w_{r>r_h}(r)dr=\dfrac{dN_1}{N}=\dfrac{\rho_1 dA_1}{\rho_1 A_1+\rho_2 A_2}=\dfrac{dA_1}{A_1+\lambda A_2}
\end{equation}

An important special case is when $\rho _2 =0$ i.e. when there is a
central hole of area $A_2=\pi r_h^2$ in the cable cross section.
In this case Eq.\ref{eq14} becomes

\begin{equation}
\label{eq16}
\begin{array}{c}
w_{r<r_h}\left( r \right)dr=\dfrac{dA_1 }{A_1}=\dfrac{dA_1 }{A-A_2 }=\dfrac{dA-dA_2 }{A-A_2 }=\\ \\
=\dfrac{2\left[{\sqrt {r_c^2 -r^2} -\sqrt {r_h^2 -r^2} } \right]}{\pi \left({r_c^2 -r_h^2 } \right)}dr
\end{array}
\end{equation}

and Eq.\ref{eq15}

\begin{equation}
\label{eq17} w_{r>r_h}=\dfrac{2\sqrt{r_c^2-r^2}}{\pi(r_c^2-r_h^2)}
\end{equation}

which gives finally the probability distribution function for a
cable with a central channel of radius $r_h $

\begin{equation}
\label{eq18} w_h\left( r \right)=\left\{ {{\begin{array}{ll}
 {\dfrac{2\left[ {\sqrt {r_c^2-r^2} -\sqrt {r_h^2 -r^2} } \right]}{\pi \left( {r_c^2
-r_h^2 } \right)} \mbox{\quad \quad for } \left| r \right| <r_h} \hfill \\
\\
{\dfrac{2 \sqrt {r_c^2-r^2} }{\pi \left( {r_c^2 -r_h^2 }
\right)}\mbox{\quad \quad for } \left| r \right| >r_h
} \hfill \\
\end{array} }} \right.
\end{equation}

\begin{figure}[tb]
\includegraphics[width=8cm,keepaspectratio]{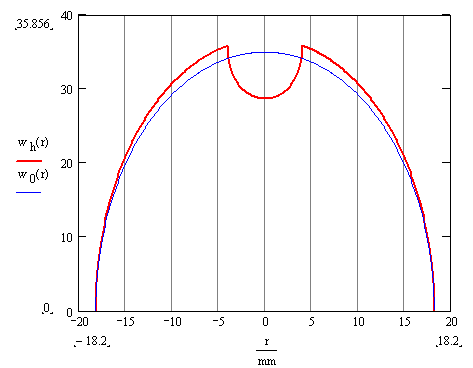}
\includegraphics[width=8cm, keepaspectratio]{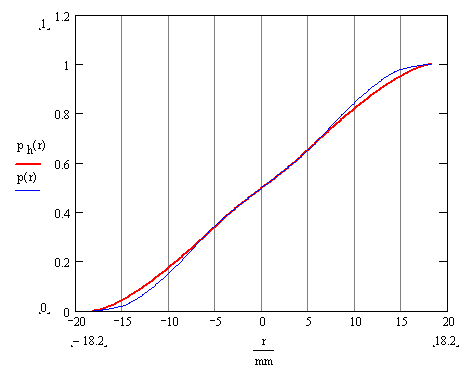}
\caption{\label{fig5}Color on line. The probability distribution function $w_h(r)$(left) and
the probability $p_h(r)$ (right) for a cable with fixed hole in the central part of
the cross-section. The blue line is the reference line for a compact cable, the probability density $w_0(r)$ and the probability $p_0(r)$.}
\end{figure}

The probability density $w_h(r)$ and the probability function $p_h(r)$
calculated  with Eq.\ref{eq18} are presented in Fig.\ref{fig5}

The example presented above shows the calculation of the probability distribution function for a cable with a fixed hole in the middle of the cross-section, a typical example of a cable with a central channel. We are going to  show
now how to calculate the probability distribution function for a
cable  with four regions, two of which being itinerant. To be more precise, the
object here is a cable in conduit conductor with wrapped
last stage petals and a central channel. As shown in
Fig.\ref{fig6} the first region is a strand region. Regions 2 and
3 are itinerant regions since they  rotate through the cross-section, 
on advancing along the cable axis, with a period given by the the 
last twist pitch of the cable. These regions are created by the upper and lower inter-petal holes. The contribution
of these two regions to the effective density is lower than that for the region one
and we call them therefore diluted (or depleted) regions. Finally, there is
a fourth region which is the central hole. This is fixed and has zero density contribution. The
following equations are  obtained by generalizing (extending) Eqs.\ref{eq13} and 14:

\begin{figure}[tb]
\centerline{\includegraphics[width=6cm,keepaspectratio]{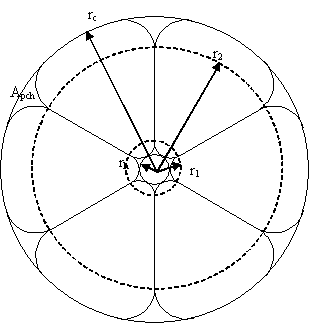}}
\caption{\label{fig6} Example of a CICC with 6 wrapped petals in
the last stage and a central hole. Observe the upper and lower
inter-petal voids. They rotate along the conductor length and are
therefore called itinerant holes (voids)as opposed to the central hole which is fixed. In this example $r_c$
=18.2~mm, $r_2$=15~mm, $r_1$ =6~mm and $r_h$ =4~mm.}
\end{figure}

\begin{equation}
\label{eq19}
\begin{array}{c}
 w\left( r \right)dr=\dfrac{\rho _1 dA_1 +\rho _2 dA_2 +\rho _3 dA_3 +\rho _4
dA_4 }{\rho _1 A_1 +\rho _2 A_2 +\rho _3 A_3 +\rho _4 A_4 }= \\ \\
 A=A_1 +A_2 +A_3 +A_4 \\
 \end{array}
\end{equation}

the first expressing the definition of the probability density function and
the second the total cross sectional area. The calculation proceeds in steps
considering different ranges for the coordinate along the field gradient,
$r$in the four regions. The four regions are

\begin{equation}
\label{eq20}
\begin{array}{l}
 \mbox{Region 1: }r_2 <\left| r \right| \\ \\
 \mbox{Region 2: }r_1 <\left| r \right|<r_2 \\ \\
 \mbox{Region 3: }r_h <\left| r \right|<r_1 \\ \\
 \mbox{Region 4: }\left| r \right|<r_h 
 \end{array}
\end{equation}

In the first region we have

\begin{equation}
\label{eq21} 
\begin{array}{c}
\left[ {w\left( r \right)dr} \right]_1 =\dfrac{\rho _2
dA_2 }{\rho _1 A_1 +\rho _2 A_2 +\rho _3 A_4 }=\\ \\
=\dfrac{\lambda _2 dA_2 }{A_1 +\lambda _2 A_2 +\lambda _3 A_3 }
\end{array}
\end{equation}

where we have defined

\begin{equation}
\label{eq22} \lambda _2 =\frac{\rho _2 }{\rho _1 };\mbox{ }\lambda
_3 =\frac{\rho _3 }{\rho _1 };\mbox{ }\lambda _4 =\frac{\rho _4
}{\rho _1 }=0
\end{equation}

where the last relation is for the central hole where no strand are present. Further we use
the following relations:

\begin{equation}
\label{eq23}
\begin{array}{c}
dA_2 =2\left(\sqrt {r_c^2 -r^2}\right)dr  \\ \\
 A_1 =\pi \left( {r_2^2 -r_1^2 } \right)  \\ \\
 A_2 =\pi \left( {r_c^2 -r_2^2 } \right)  \\ \\
 A_3 =\pi \left( {r_1^2 -r_h^2 } \right)  \\ \\
 A_4 =\pi r_h^2 \\ \\
\end{array}
\end{equation}

The dilution factors for the two itinerant regions are

\begin{equation}
\label{eq24}
\begin{array}{c} 
\lambda _2 =\dfrac{\rho _2 }{\rho _1 }=\dfrac{1}{\rho
_1 }\dfrac{N_2 }{A_2 }=\dfrac{1}{\rho _1 }\dfrac{\rho _1 \left( {A_2
-6A_{pch} } \right)}{A_2 }=\\ \\ 
=1-\dfrac{6A_{pch} }{A_2 }
\end{array}
\end{equation}

where $A_{pch}$ is the area of the inter petal void and the factor
6 comes from the fact that there are six inter petal cable
voids. Similarly, if $a_{pch}$ is the area of the lower
inter-petal void we have

\begin{equation}
\label{eq25} \lambda _3 =\frac{\rho _3 }{\rho _1
}=1-\frac{6a_{pch} }{A_3 }
\end{equation}

In general we observe that the dilution factor for the itinerant regions
is always of the form

\begin{equation}
\label{eq26} \lambda _i =1-\frac{A_{i,void} }{A_i }
\end{equation}

and are intermediate between full stranded regions where $\lambda
=1$ and fixed hole regions with $\lambda =0$. The name diluted
regions comes from this fact.

Using Eqs.\ref{eq24} and \ref{eq25}, the denominator in Eq.\ref{eq21} can be
expressed as

\begin{eqnarray}
\label{eq27}
A_1 +\lambda _2 A_2 +\lambda _3 A_3 = \\ \nonumber
 =A_1 +\left( {1-\dfrac{A_{2,void} }{A_2}} \right)A_2 +\left( {1-\dfrac{A_{3,void} }{A_3 }} \right)A_3 = \\ \nonumber
 =A_1 +A_2 +A_3 -A_{2,void} -A_{3,void} = \\ \nonumber
 =A-A_{2,void} -A_{3,void} -A_4
 \end{eqnarray}

Finally we get the following expression for the probability
density function in Region 1

\begin{equation}
\label{eq28} \left[ {w\left( r \right)} \right]_1 =\frac{2\sqrt
{r_c^2 -r^2} \left( {1-\dfrac{A_{2,void} }{A_2 }}
\right)}{A-A_{2,void} -A_{3,void} -A_4 }
\end{equation}

In the second region we have

\begin{equation}
\label{eq29}
\begin{array}{l}
 \left[ {w\left( r \right)dr} \right]_2 =\dfrac{\rho _1 dA_1 +\rho _2 dA_2
}{A_1 +\lambda_2 A_2 +\lambda_3 A_3 }=\dfrac{dA_1 +\lambda _2
dA_2 }{\rho _1 A_1 +\rho _2 A_2 +\rho _3 A_3 }= \\ \\=\dfrac{dA_1 +\lambda _2 dA_2 }{A-A_{2,void} -A_{3,void} -A_4 } \\
 \end{array}
\end{equation}

where we have used Eq.\ref{eq27}. Now, observe that

\begin{eqnarray}
\label{eq30}
 dA_1 =2\sqrt {r_2^2 -r^2} dr \\ \nonumber
 dA_2 =2\left[ {\sqrt {r_c^2 -r^2} -\sqrt {r_2^2 -r^2} } \right]dr
\end{eqnarray}

and from  Eq.\ref{eq29} using Eq.\ref{eq26} we obtain the probability density function in the region 2 as

\begin{widetext}
\begin{equation}
\label{eq31} \left[ {w\left( r \right)} \right]_2 =\dfrac{2\sqrt
{r_2^2 -r^2} +2\left( {1-\dfrac{A_{2,void} }{A_2 }} \right)\left(
{\sqrt {r_c^2 -r^2} -\sqrt {r_2^2 -r^2} }
\right)}{A-A_{2,void}-A_{3,void} -A_4 }
\end{equation}
\end{widetext}

In the third region we have $r_h <\left| r \right|<r_1 $ and
according to the general definition we have

\begin{equation}
\label{eq32}
\begin{array}{c}
 \left[ {w\left( r \right)dr} \right]_3 =\dfrac{\rho _1 dA_1 +\rho _2 dA_2
+\rho _3 dA_3 }{\rho _1 A_1 +\rho _2 A_2 +\rho _3 A_3}=\\ \\ 
=\dfrac{dA_1 +\lambda_2 dA_2 +\lambda _3 dA_3 }{ A_1 +\lambda_2 A_2 +\lambda_3 A_3 }= \\ \\
 =\dfrac{dA_1 +\lambda _2 dA_2 +\lambda _3 dA_3 }{A-A_{2,void} -A_{3.void}-A_4 } \\
 \end{array}
\end{equation}

where we have first divided the nominator and denominator by $\rho_1$ and then have used Eq.\ref{eq27}. Now we write again the expressions for the elemental areas of interest which in the third region are:

\begin{equation}
\label{eq33}
\begin{array}{c}
 dA_1 =2\left( {\sqrt {r_2^2 -r^2} -\sqrt {r_1^2 -r^2} } \right)dr \\ \\
 dA_2 =2\left( {\sqrt {r_c^2 -r^2} -\sqrt {r_2^2 -r^2} } \right)dr \\ \\
 dA_3 =2\sqrt {r_1^2 -r^2} dr \\
 \end{array}
\end{equation}

and after substituting in Eq.\ref{eq32}and simplifying $dr$ we get

\begin{widetext}
\begin{equation}
\label{eq34}
\left[ {w\left( r \right)} \right]_3 = \dfrac{2\left({\sqrt {r_2^2
-r^2} -\sqrt {r_1^2 -r^2} } \right)+2\left( {1-\dfrac{A_{2,void}
}{A_2 }} \right)\left( {\sqrt {r_c^2 -r^2} -\sqrt {r_2^2 -r^2} }
\right)+2\left( {1-\dfrac{A_{3,void} }{A_3 }} \right)\sqrt
{r_1^2 -r^2}}{A-A_{2,void} -A_{3,void} -A_4 }
\end{equation}
\end{widetext}

Finally, we have the fourth region with $\left| r \right|<r_h $.
The probability distribution is

\begin{equation}
\label{eq35}
\begin{array}{c}
\left[ {w\left( r \right)dr} \right]_4 =\dfrac{\rho
_1 dA_1 +\rho _2 dA_2 +\rho _3 dA_3 }{\rho _1 A_1 +\rho _2 A_2+\rho _3 A_3 }=\\ \\
=\dfrac{dA_1 +\lambda_2 dA_2 +\lambda_3 dA_3}{A-A_{2,void} -A_{3,void} -A_4 } \\
\end{array}
\end{equation}

with the new elemental areas for this region

\begin{equation}
\label{eq36}
\begin{array}{l}
 dA_1 =2\left( {\sqrt {r_2^2 -r^2} -\sqrt {r_1^2 -r^2} } \right)dr \\
 dA_2 =2\left( {\sqrt {r_c^2 -r^2} -\sqrt {r_2^2 -r^2} } \right)dr \\
 dA_3 =2\left( {\sqrt {r_1^2 -r^2} -\sqrt {r_h^2 -r^2} } \right)dr \\ \\
\end{array}
\end{equation}

After substitution in Eq.\ref{eq35} and simplifying $dr$ on both sides of the equation we obtain

\begin{widetext}
\begin{equation}
\label{eq37}
\begin{array}{c}
 \left[ {w\left( r \right)} \right]_4 =\\ \\
 =\dfrac{2\left( {\sqrt {r_2^2 -r^2}
-\sqrt {r_1^2 -r^2} } \right)+2\left( {1-\dfrac{A_{2,void} }{A_2
}} \right)\left( {\sqrt {r_c^2 -r^2} -\sqrt {r_2^2 -r^2} }
\right)+2\left({1-\dfrac{A_{3,void} }{A_3 }} \right)\left(
{\sqrt{r_1^2-r^2} -\sqrt {r_h^2 -r^2} } \right)}{A-A_{2,void} -A_{3,void} -A_4 } \\
\end{array}
\end{equation}
\end{widetext}

\begin{figure}[tb]
\centerline{\includegraphics[width=8cm,keepaspectratio]{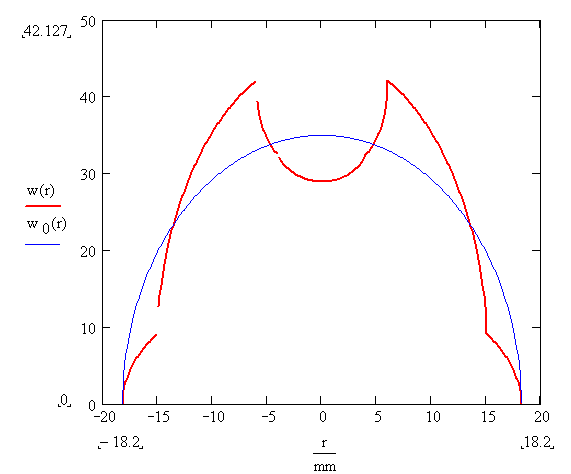}}
\caption{\label{fig7} Color online. Probability density function for the CICC
calculated with the relations developed in this paper. For
reference also the probability density for a simple connected
cable is shown.}
\end{figure}

The final result for the probability density function is obtained by putting together the four pieces calculated before, Eq.\ref{eq28}, Eq.\ref{eq31}, Eq.\ref{eq34} and Eq.\ref{eq37}, in the form of a piecewise continuous function which is plotted in Fig.\ref{fig7} 

\begin{equation}
\label{eq38} w\left( r \right)=\left\{ {{\begin{array}{*{20}l}
 {\left[ {w\left( r \right)} \right]_1 \mbox{ for }r_2 <\left| r \right|}\hfill \\ \\
 {\left[ {w\left( r \right)} \right]_2 \mbox{ for }r_1 <\left| r \right|<r_2} \hfill \\ \\
 {\left[ {w\left( r \right)} \right]_3 \mbox{ for }r_h <\left| r \right|<r_1} \hfill \\ \\
 {\left[ {w\left( r \right)} \right]_4 \mbox{ for }\left| r \right|<r_h }\hfill \\
\end{array} }} \right.
\end{equation}

\section{A cable-in-conduit conductor with segregated copper}
\begin{figure}[b]
  \includegraphics[width=6cm]{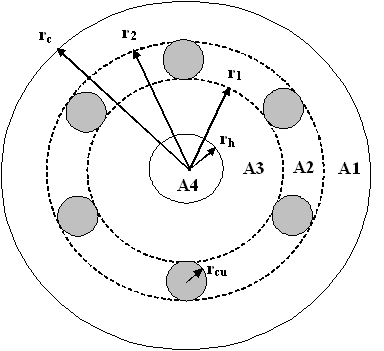}
  \caption{Geometry data for the advanced Nb3Sn cable with six itinerant copper
  islands and a central hole}\label{fig8}
\end{figure}

Next we illustrate the general method to calculate the geometrical
probability function for a cable with segregated, discrete copper.
Discrete means that the copper wires, which replace some of the
superconducting strands, are not dispersed uniformly in the cable
cross-section but grouped together in what appears as islands in
the cable structure. The geometry, presented in Fig.\ref{fig8} is
borrowed from the new advanced $Nb_3Sn$ conductor \cite{ref5}, a
test conductor developed and tested at the SULTAN
facility. As can be seen, in the central cable region there are
six round copper islands each being a braid of 27 copper wires.
The islands are itinerant i.e. rotate in the cross-section as one
advance along the cable. As before we define different regions
starting from the cable jacket with a radius $r_c$. The region 1
is the reference region and we take $\lambda_1=1$. The second
region is a diluted region due to the circular movement of the
copper islands and the dilution factor will be calculated later.
The third region is as the first region and we take $\lambda_3=1$.
Finally, the fourth region is a hole with radius $r_h$ and
therefore we have here $\rho_4=0$ and $\lambda_4=0$.

In the first region, $r_2<|r|<r_c$ and we have therefore

\begin{equation}
\begin{array}{c}
\label{eq39} 
[w(r)]_1=\dfrac{\rho_1dA_1}{\rho_1 A_1+\rho_2 A_2 +
\rho_3 A_3}=\\ \\
=\dfrac{dA_1}{A_1 +\lambda_2 A_2 +A_3} \\
\end{array}
\end{equation}

Further we have $dA_1=2\left(\sqrt{r_c^2-r^2}\right)dr$ and

\begin{equation}
\begin{array}{c}
\label{eq40} 
A_1+\lambda_2A_2+A_3=A_1+(A_2-A_{2,void})+A_3= \\ \\
=A_1+A_2+A_3-A_{2,void}=A-A_{2,void}-A_4 \\
\end{array}
\end{equation}

where $A=\pi r_c^2$ is the total cable area, $A_4=\pi r_h^2$ the
hole area, $A_{2,void}=6\pi r_{Cu}^2$ and we used the fact that $A_1+A_2+A_3=A-A_4$. Substituting all these in
Eq.\ref{eq39} we get

\begin{equation}
\label{eq41}
[w(r)]_1=\dfrac{ 2\sqrt{r_c^2-r^2} }{A-A_{2,void}-A_4}=\dfrac{ 2\sqrt{r_c^2-r^2} }{A_{eff}}
\end{equation}

where in order to simplify the notation we defined $A_{eff}=A-A_{2,void}-A_4$.

In the second region, $r_1<|r|<r_2$ the probability density is

\begin{equation}
\label{eq42}
    [w(r)]_2=\dfrac{\rho_1 dA_1+\rho_2 dA_2}{\rho_1 A_1+\rho_2 A_2+\rho_3
    A_3}=\dfrac{dA_1+\lambda_2 dA_2}{A_{eff}}
\end{equation}

and with $dA_1=2\left[\sqrt{r_c^2-r^2}-\sqrt{r_2^2-r^2}\right]dr$ and $dA_2=2\sqrt{r_2^2-r^2}dr$
we obtain

\begin{equation}
\label{eq43}
    [w(r)]_2=\dfrac{ 2\sqrt{r_c^2-r^2}+2\lambda_2 \sqrt{r_2^2-r^2} }{A_{eff}}
\end{equation}

For the other regions the calculation follows the same scheme and we give here only the
final result. For the third and the fourth region we get

\begin{widetext}
\begin{equation}\label{eq44}
\begin{array}{c}
    [w(r)]_3=\dfrac{ 2\left[\sqrt{r_c^2-r^2}-\sqrt{r_2^2-r^2}\right]+2\lambda_2\left[\sqrt{r_2^2-r^2}
-\sqrt{r_1^2-r^2}\right]+2\sqrt{r_1^2-r^2}}{A_{eff}} \\ \\
		\left[w(r)\right]_4=\dfrac{ 2\left[\sqrt{r_c^2-r^2}-\sqrt{r_2^2-r^2}\right]+2\lambda_2\left[\sqrt{r_2^2-r^2}
-\sqrt{r_1^2-r^2}\right]+2\left[\sqrt{r_1^2-r^2}-\sqrt{r_h^2-r^2}\right]}{A_{eff}}
\end{array}    
\end{equation}
\end{widetext}

The dilution factor for the second region containing the copper islands is

\begin{equation}\label{eq45}
    \lambda_2=1-\dfrac{A_{2,void}}{A_2}=1-\dfrac{6\pi r_{Cu}^2}{\pi (r_2^2-r_1^2)}
\end{equation}

since the void area is equal to the area of the six copper islands.

\begin{figure}[tb]
  \includegraphics[width=8cm]{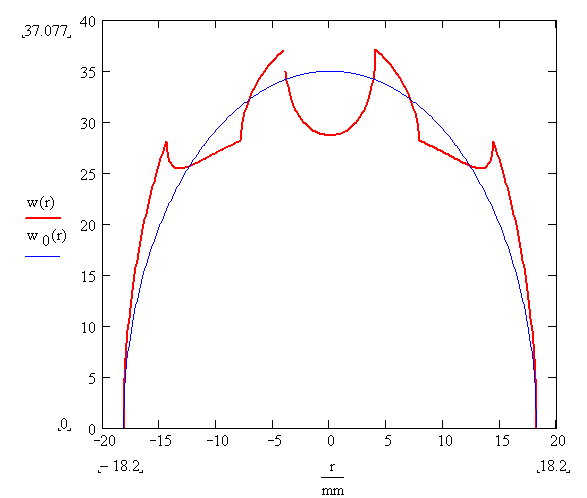}\\
  \caption{Color online. The probability distribution function for the cable with six copper islands
  and a central hole}\label{fig9}
\end{figure}

The final solution is then of the same form as Eq.\ref{eq38} i.e. a piece-wise continuous function.
The result of numerical calculation is shown in Fig.\ref{fig9} and
corresponds to the following set of parameters: $r_h$=4~mm,
$r_{Cu}$=3.25~mm, $r_1$=7.85~mm, $r_2$=14.35~mm and $r_c$=18.2~mm.

\section{Conclusions}
A method and general relations for calculating the probability
distribution function for cable-in-conduit conductors with
different geometry have been presented. The probability distribution
function is an essential ingredient in the calculation of the
average electric field of a cable carrying transport current and
exposed to an external uniform magnetic field. This is the only
calculation which can be compared with the measured electric field and is an important element in the DC
characterization of a cable. 

The method was illustrated for two favorite cable geometries with multiple-connected topology. 
The results of the calculation for a cable with
six petals and a central hole and for a cable with six
segregated copper islands show that neglecting these geometric
aspects could led to large errors in evaluating the average
electric field. The central hole is seen to induce a depletion of
the probability distribution function in the central region of the
cable. Wrapping the last stage results in a depletion close to the
jacket and an enhancement close to and around the central hole as
illustrated in Fig.\ref{fig7}. It is worth mentioning that the
depletion in this case occurs at and close to the peak field
position. The copper islands effect, Fig.\ref{fig9} is also a
depletion but strongly localized around the annulus where the
copper segregation is present. 
The calculation method presented here can be use for almost any type of cable satisfying the ergodic principle.

\end{document}